\documentclass{article}[12pts]
\title{Duality and even number spin-correlation functions in the Two dimensional square lattice Ising model}
\author{Ranjan Kumar Ghosh\thanks{\normalsize e-mail address:- rkg\_1978@yahoo.com}\\Bidhannagar College\\EB-2,Salt Lake City,\\Calcutta-700 064, INDIA}
\date{March 20, 2007}
\begin{document}
\maketitle
\begin{abstract}\normalsize
The Kramers-Wannier duality is shown to hold for all the even number spin correlation functions of the   two dimensional square lattice Ising model in the sense that the high temperature $(T> T_{c})$ expressions for these correlation functions are transformed into the low temperature $(T< T_{c})$ expressions under this duality transformation.
\end{abstract}

	The two dimensional Ising model remains one of the very few models for which many properties can be calculated exactly$ ^{[1,3,4,5,6,7,8,9]}$. Among these are the partition function and in principle all the multi-spin correlation functions for the case of the square lattice (isotropic ) model. At least the algorithm for computing all the multispin correlation functions is known$^{[10,11}]$.Quite a few of the two-spin correlation functions have been computed in terms of the complete elliptic integrals of the first and the second kind. It is also well known that the expression for the diagonal two-spin correlation functions for the anisotropic model have the same expression in terms of the complete elliptic integrals as that of the isotropic model. These correlation functions are known to satisfy a $Painleve' VI$ nonlinear ODE with a proper definition of the variables$^{[13]}$. All these things perhaps make it arguably the most widely studied model of ferromagnetism and antiferromagnetism\\

	Even though the model is so widely studied, new properties of this model are being discovered from time to time. It is the purpose of this communication to discuss one of such properties of the model.\\

The very first sign that the model shows a phase transition from a disordered state to a ordered state was discovered by Kramers and Wannier$^{[2]}$ by considering what is known as the duality property of the partition function of the model. We give here the Hamiltonian for the model which is\begin{equation}
H=-J\sum_{n.n}\left[\sigma_{i,j}\sigma_{i,j +1}+\sigma_{i,j}\sigma_{i +1,j}\right]\end{equation}

	where $\sigma_{i,j}=\pm {1}$, the products in the summation include only the nearest neighbours and $i$ and $j$ denote the positions of the spin on the lattice. The  interaction strength $J$ is positive for the ferromagnetic model and negative for the antiferromagnetic model.\\

	The partition function of the model can then be written as
\begin{equation}Z=\sum_{\sigma=\pm 1}\exp^{-\beta H}\end{equation}
where$\beta=\left(k_{B}T\right)^{-1}$,
and any multi-spin correlation function can be written as
\begin{equation}<\sigma_{i,j}\sigma_{k,l} \cdots>=\frac{1}{Z}\sum_{\sigma=\pm 1}\sigma_{i,j}\sigma_{k,l}\cdots \exp^{-\beta H}\end{equation}

	The Kramers-Wannier duality can be understood by considering the high temperature and low temperature expansions of the partition function in the appropriate expansion variables. It is seen that each of the configurations of the high temperature expansion side has a counterpart in the low temperature expansion, and thus there is a one to one correspondence between the two phases of the theory. Kramers and Wannier$^{[2]}$ used this high temperature -low temperature duality to find the critical temperature for the model, the temperature at which the model undergoes a phase transition. According to their calculations, the critical temperature is given by\begin{equation}(sinh2\beta J)^{2}=1\end{equation}\\

Later, in a celebrated paper, Onsager$^{[3]}$ calculated the exact partition function by considering the transfer matrix for this model, which proved the correctness of Kramers-Wannier formula. The next quantity in line was the expctation value of a single spin, or in other words the spontaneous magnetization of the model, the form for which was given by Onsager and proved by Yang$^{[4]}$. Since then there were many numerical calculations of the multispin expectation values which finally resulted in writing down some of the two spin correlation functions of the model in terms of the complete elliptic integrals showing a beautiful structure of these functions when exact expressions are written for these correlation functions in terms of these integrals$^{[7,8,9]}$. It was also clear that all the two-spin correlation functions can indeed be written in terms of the complete elliptic integrals.The two-spin correlation functions were also shown to satisfy certain quadratic difference equations that were derived by Perk$^{[12]}$. However it was not very clear how these quantities are related in the two phases of the model.\\

In the meantime much work was done to understand the diagonal two-spin correlation functions and Jimbo and Miwa developed an equation that these correlation functions satisfy.General series solutions to this equation were dicussed in ref.$^{[14,15,18]}$.\\

For the exact expressions for some of the correlation functions of this type the reader is referred to ref.[7,17]. We can take any of these expressions as a function of the variable $k=\left(\sinh 2\beta J\right)^2$ or $1/k=\left(\sinh 2\beta J\right)^2$ for $T>T_{c}$ or $T<T_{c}$ respectively and consider the following non linear transformation on it.\begin{equation}k\rightarrow \frac{1}{k}\end{equation}\begin{equation}K\left(k\right)\rightarrow k K\left(k\right)\end{equation}\begin{equation}E\left(k\right)\rightarrow\frac{E\left(k\right)+\left(k^{2}-1\right)K\left(k\right)}{k}\end{equation}

	The last two transformation equations are just seen to be the correct transformation rules for the complete elliptic integrals from their transformation properties. Under these transformations simultaneously, the high temperature $\left(T>T_{c}\right)$ expressions are seen to transform into the low temperature $\left(T<T_{c}\right)$ expressions and vice versa. This is true for all of the known diagonal two-spin correlation functions. However there is nothing special about these correlation functions and this property must be true for all the spin correlation functions that can be written in terms of $k$ and the complete elliptic integrals $K\left(k\right)$ and $E\left(k\right)$ and therefore this property must be true for all the even number spin correlation functions, which share this dependence on $k$ and the complete elliptic integrals. It is essentially seen to be the Kramers-Wannier duality at work again.\\

We now proceed to give a proof of the statement made above, from the Montroll, Potts and Ward's formulation$^{[5]}$ of even number spin correlation problem of the isotropic Ising model in terms of Pfaffians. In our notation, we will follow their paper as closely as possible.\\

	At first we consider the case of the rectangular lattice Ising model. Its partition function is given by\begin{equation} Z=\sum_{\sigma=\pm1}\prod_{n.n} \exp\left(\beta J_{1}\sigma_{i,j}\sigma_{i,j+1}+\beta J_{2}\sigma_{i,j}\sigma_{i+1,i}\right)\end{equation}

	where $\sigma_{i,j}=\pm1$ is the state of the spin at the $\left(i,j\right)$ site of the lattice.$i$ increases in one direction and $j$ in the perpendicular direction. The product includes all the nearest neighbours and the sum is over all the spins of the lattice. This, for the isotropic case $J_{1}=J_{2}$ reduces to expression(1). The transition temperature of this model is given by\begin{equation}\sinh 2\beta J_{1} \sinh 2\beta J_{2}=1\end{equation}
	
	 We introduce the variables \begin{equation}z_{1}=\tanh \beta J_{1}\ \ \ z_{2}=\tanh \beta J_{2}\end{equation}
With the help of these variables the partition function can be written as\begin{eqnarray}Z =\left(\cosh \beta J_{1} \cosh \beta J_{2}\right)^N \sum_{\sigma=\pm 1} \prod_{n.n} \left(1+z_{1}\sigma_{i,j}\sigma_{i,j+1}\right)\nonumber\\\times\left(1+z_{2}\sigma_{i,j}\sigma_{i+1,j}\right)\end{eqnarray}

	where $N$ is the total number of spins on the lattice. The even number spin correlation functions can be formally written down as\begin{eqnarray}<\sigma_{0,0}\sigma_{m,n}\sigma_{p,q}\cdots>=Z^{-1}\left(\cosh \beta J_{1} \cosh \beta J_{2}\right)^N \times\sum_{\sigma=\pm 1}\sigma_{0,0}\sigma_{m,n}\sigma_{p,q}\cdots\nonumber\\ \times\prod_{n.n}\left(1+z_{1}\sigma_{i,j}\sigma_{i,j+1}\right)\left(1+z_{2}\sigma_{i,j}\sigma_{i+1,j}\right)\end{eqnarray}

	 		where the product of spins on the l.h.s. contains an even number of spins.Following ref.[5] we can connect pairs of spins by (non overlapping) continuous line segments and we can introduce the factors of $\sigma_{j,k}^2=1$ along those links and then arrange the resulting expression in such a way that it is related to the original partition function with a perturbed interaction strength along those links. One can then relate this expression to the Pfaffians by the usual correspondence.\\

As discussed by the authors of ref.[5], this results in expressions which contain terms and integrals which can be related to the complete elliptic integrals of the first and the second kind in the case when we have the interaction strengths equal along both the vertical and horizontal directions, i.e, we have a square lattice.\\

	If we take the argument of these integrals to be the variable $k$ in one phase of the system then the expressions for the other phase are obtained by simply replacing $k$ by $\frac{1}{k}$ everywhere in the expressions because \textit {there is only one expression for the correlation functions in terms of the Pfaffians, the one given by equation(12)}.To make the argument clearer, we take one of the simplest cases, the case of $<\sigma_{0,0}\sigma_{0,1}>$ which is also one of the cases discussed in ref[5].The generalization of the logic is obvious and therefore this case illustrates the properties of the general even number spin correlations quite simply. In the notation of ref.[5] this two-spin correlation function is given by 
	\begin{eqnarray} <\sigma_{0,0}\sigma_{0,1}>=\frac{1}{Z} z_{1}\left(\cosh \beta J_{1} \cosh \beta J_{2}\right)^{N}\sum_{\sigma=\pm 1}\left(1+z_{1}^{-1}\sigma_{0,0}\sigma_{0,1}\right) \nonumber\\
\times	\prod_{n.n}'\left(1+z_{1} \sigma_ {i,j}\sigma_{i,j+1}\right)\left(1+z_{2}\sigma_{i,j}\sigma_{i+1,j}\right)\end{eqnarray}

	where the prime shows that the factors containing the product of spins which have occured previously in the expression have to be omitted in the later product.

	 after going through some arguments this expression can be written as the product of the Pfaffian of\begin{equation}\left(z_{1}^{-1}-z_{1}\right)\left[\begin{array}{cc}0 & 1\\
																																	-1 & 0\end{array}\right]\end{equation}
and the Pfaffian of
\begin{equation}\left[\begin{array}{cc}0 & -\left(z_{1}^{-1}-z_{1}\right)^{-1}+\left[0,1\right]_{RL}\\
                                     \left(z_{1}^{-1}-z_{1}\right)^{-1}-\left[0,1\right]_{RL} &0\end{array} \right]\end{equation}     																																	and $z_{1}$, where $\left[0,1\right]_{RL}$ is one of the elements of the matrix defined in the ref.[5].
 The net result of all this is that\begin{equation}<\sigma_{0,0}\sigma_{0,1}>=\pm\left(z_{1}-\left(1-z_{1}^2\right)\left[0,1\right]_{RL}\right)\end{equation} \\

 This for the isotropic case when$z_{1}=z_{2}=z=\tanh \beta J$ yields                                   
	 \begin{equation}<\sigma_{0,0}\sigma_{0,1}>=\pm \left(z-\left(1-z^{2}\right)\left[0,1\right]_{RL}\right)\end{equation}

However the $z$ as well as the element $\left[0,1\right]_{RL}$ can be written in terms of the variables $k=\left(sinh 2\beta J\right)^{2}$ or the low temperature expansion variable for which $\frac{1}{k}=\left(sinh 2\beta J\right)^{2}$. As is clear the double integral expressions for the matrix elements involved can be written in term of complete elliptic integrals with the modulus $k$ in the high temperature phase and modulus $\frac{1}{k}$ in the low temperature phase.
One therefore just has to look for the transformation properties of the complete elliptic integrals under the transformation $$k\rightarrow\frac{1}{k}$$ and these are seen to be given by the equations (6) and (7)\footnote{\ see for example ref.[16]}.\\

	This therefore has provided us with a very important tool for relating the high temperature expressions to the low temperature expressions and vice versa. In general this means that we actually only need to know the $T<T_{c}$ expressions for any arbitrary correlation function because the correlation functions involving an odd number of spins vanishes in the high temperature phase. Thus just the knowledge of $T<T_{c}$ expressions provide a complete description of the system.\\

	As a special case of these relations, in the $Painleve'VI$ equation of Jimbo and Miwa, the solutions for integral values of $n$ in the two phases are related to each other by the above mentioned transformations. But,of course, this remark is applicable only for the exact solutions and is not applicable to the series solutions of this equation term by term.\\

	We have thus learnt that the Kramers-Wannier duality has a greater validity than being applicable only to the partition function of the square lattice Ising model, it extends to all the even number spin-correlation functions of the model.

\end{document}